\documentclass[10pt, conference, letterpaper]{IEEEtran}
\IEEEoverridecommandlockouts
\usepackage{cite}
\usepackage{graphicx}
\usepackage{subfig}
\usepackage{textcomp}
\usepackage{xcolor,bm}
\def\BibTeX{{\rm B\kern-.05em{\sc i\kern-.025em b}\kern-.08em
    T\kern-.1667em\lower.7ex\hbox{E}\kern-.125emX}}
    
\usepackage{times,graphics,mathptm,epsfig,amsmath,amsfonts, xspace,endnotes,pifont,multirow,rotating,listings,amssymb,algorithmic,color,caption,nicefrac,adjustbox,todonotes,tabularx,mathtools, algorithmic,algorithm}
\usepackage{tikz}
\usepackage{textcomp}
\usepackage{lipsum}

\newcommand\copyrighttext{%
  \footnotesize \textcopyright 2020 IEEE. Personal use of this material is permitted.
  Permission from IEEE must be obtained for all other uses, in any current or future 
  media, including reprinting/republishing this material for advertising or promotional 
  purposes, creating new collective works, for resale or redistribution to servers or 
  lists, or reuse of any copyrighted component of this work in other works. 
  DOI: 10.1109/ICC40277.2020.9149175
  }
\newcommand\copyrightnotice{%
\begin{tikzpicture}[remember picture,overlay]
\node[anchor=south,yshift=10pt] at (current page.south) {\fbox{\parbox{\dimexpr\textwidth-\fboxsep-\fboxrule\relax}{\copyrighttext}}};
\end{tikzpicture}%
}

\begin{document}

\title{Joint Satellite Gateway Placement and Routing for Integrated Satellite-Terrestrial Networks}


\makeatletter
\newcommand{\linebreakand}{%
  \end{@IEEEauthorhalign}
  \hfill\mbox{}\par
  \mbox{}\hfill\begin{@IEEEauthorhalign}
}
\makeatother

\author{Nariman Torkzaban$^{\dagger}$, Anousheh Gholami$^{\dagger}$, Chrysa Papagianni$^{\ddagger}$, and John S. Baras$^{\dagger}$\\
\tt\small $^{\dagger}$Department of Electrical and Computer Engineering\\
Institute for Systems Research\\
University of Maryland, College Park, MD 20742, USA\\
Email: \tt\small \{ nariamnt | anousheh | baras\} @umd.edu \\
\tt\small $^{\ddagger}$ Nokia Bell Labs, Antwerp, Belgium\\
Email: chrysa.papagianni@nokia-bell-labs.com
}

\maketitle
\copyrightnotice

\begin{abstract}
With the increasing attention to the integrated satellite-terrestrial networks (ISTNs), the satellite gateway placement problem becomes of paramount importance. The resulting network performance may vary depending on the different design strategies.
In this paper a joint satellite gateway placement and routing strategy for the terrestrial network is proposed to minimize the overall cost of gateway deployment and traffic routing, while adhering to the average delay requirement for traffic demands. Although traffic routing and gateway placement can be solved independently, the dependence between the routing decisions for different demands makes it more realistic to solve an aggregated model instead. 
We develop a mixed integer linear program (MILP) formulation for the problem. We relax the integrality constraints to achieve a linear program (LP) which reduces time-complexity at the expense of a sub-optimal solution. 
We further propose a variant of the proposed model to balance the load between the selected gateways.
\end{abstract}

\IEEEpeerreviewmaketitle
\section{Introduction}
Over the past years, the integration of satellite communications with current and emerging terrestrial networks (e.g., 5G mobile networks) is gaining attention, given the growing data traffic volume which is predicted to increase by over 10,000 times in the next 20 years \cite{8438267}. The trend is also supported by national directives and initiatives to support broadband connectivity in rural and remote areas, as it is considered a crucial factor for economic growth. Due to their large footprint, satellites can complement and extend terrestrial networks, both in densely populated areas and in rural zones, and provide reliable mission critical services. Standardization bodies as 3GPP, ETSI \cite{doi:10.1002/sat.1292} and ITU \cite{8438267} also recognize and promote integrated and/or hybrid satellite and terrestrial networks.

The integrated satellite-terrestrial networks (ISTNs) can be a cornerstone to the realization of a heterogeneous global system to enhance the end-user experience \cite{artiga2016terrestrial}. An ISTN, as depicted in Fig.~\ref{fig:map}, is composed of satellites organized in a constellation that support routing, adaptive access control, and spot-beam management \cite{vasavada2016architectures}, whereas the  terrestrial optical network consists of ground stations (gateways), switches, and servers. 
Delay-sensitive data services are more suitable to transport in the low earth orbit (LEO) satellite networks, which provide inherent advantages in power consumption, pole coverage, latency, and lower cost compared with the geostationary earth orbit (GEO) satellite networks and the medium earth orbit (MEO) satellite networks \cite{guo19}. 

\begin{figure}[t]
\begin{center}
\includegraphics[width=0.45\textwidth]{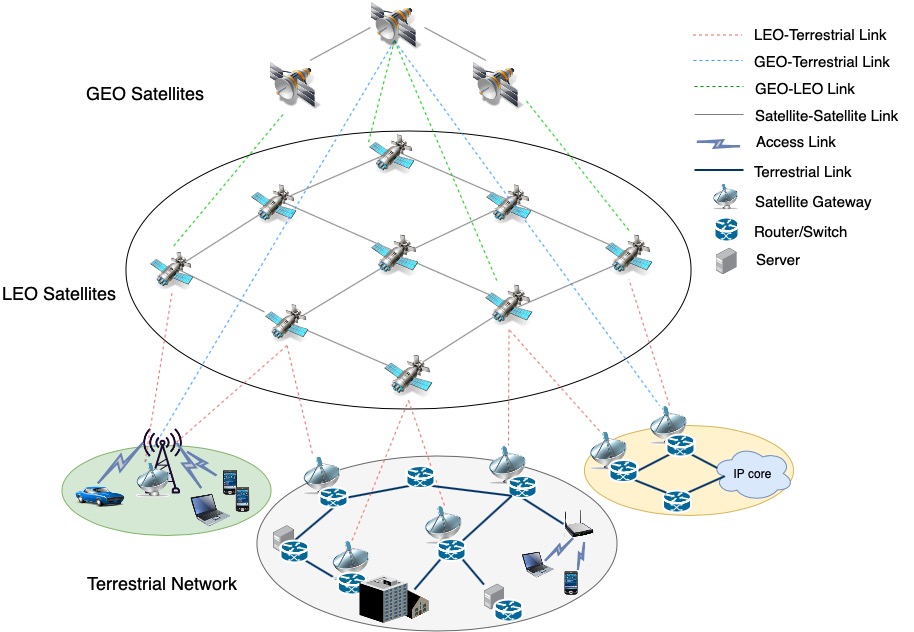}
\caption{Example of an integrated satellite-terrestrial network.}
\label{fig:map}
\end{center}
\end{figure}

However, there are multiple challenges associated with the ISTNs. Satellite networks and terrestrial networks are widely heterogeneous, supported by distinct control and management frameworks, thus their integration to date was based on proprietary and custom solutions.  Additionally, in the integrated network, as there are multiple available paths along which data traffic can be routed, optimized path-selection is important to satisfy the Quality of Service (QoS) requirements of the traffic flows and improve the utilization of network resources \cite{jia2016broadband}. Existing satellite networks employ a decentralized management architecture, scheduled link allocation and static routing strategies, which make it difficult to support flexible traffic scheduling adapting to the changes in traffic demands \cite{service}. Complex handover mechanisms should be in place at the gateway nodes, while their high-power consumption requirements need to be also taken into consideration.

Nowadays the inclusion of plug-and-play SatCom solutions are investigated, primarily in the context of 5G. Exploiting software-defined networking (SDN)\cite{zhang2017energy} and network function virtualization (NFV), ISTNs can be programmed and network behavior can be centrally controlled using open APIs, thus supporting a more agile integration of terrestrial (wired and wireless) and satellite domains interworking with each other to deliver end-to-end services. Based on the extra degrees of freedom in ISTN deployment, brought forth by the softwarization and programmability of the network, the satellite gateway placement problem is of paramount importance for the space-ground integrated network. The problem entails the selection of an optimal subset of the terrestrial nodes for hosting the satellite gateways while satisfying a group of design requirements. Various strategic objectives can be pursued when deciding the optimal placement of the satellite gateways including but not limited to cost reduction, latency minimization, reliability assurance, etc. \cite{GW}, \cite{GWC}, \cite{taghlid}.

 In this paper, we propose a method for the cost-optimal deployment of satellite gateways on the terrestrial nodes. Particularly, we aim to minimize the overall cost of gateway deployment and traffic routing, while satisfying latency requirements. To this end, we formulate the problem as a mixed-integer linear program (MILP), with latency bounds as hard constraints. We derive an approximation method from our MILP model to significantly reduce the time-complexity of the solution, at the expense of sub-optimal gateway placements, and investigate the corresponding trade-off. Furthermore, in order to reduce latency and processing power at the gateways, we impose a (varying) upper bound on the load that can be supported by each gateway.

It is important to note that, traffic routing and facility placement
are usually solved as different problems, but assigning
a demand point to a facility without considering the other
demand points might not be realistic [9]. Given the significant
interrelation of the two problems we develop and solve a single
aggregated model instead of solving the two problems in a
sequential manner.

The remainder of the paper is organized as follows. Section~\ref{sec:desc} describes the problem and the network model. In Section~\ref{sec:problem} we introduce the MILP formulation and its LP-based approximation followed by a variant of our model with load minimization. We present our evaluation results in section \ref{sec:evaluation}, and provide the overview of the related works in section \ref{sec:relatedwork}. Finally, in Section \ref{sec:conclusions}, we highlight our conclusions and discuss the directions for our 
future work.
\section{Network Model and Problem Description} 
\label{sec:desc}
The ISTN network under our study is depicted in Fig. \ref{fig:map}.We model the terrestrial network as an undirected $G = (V, E)$ graph where $(u,v) \in E $ if there is a link between the nodes $u,v \in V$. Let $J\subseteq V $ be the set of all potential nodes for gateway placement and $I\subseteq V$ be the set of all demand points. We note that the sets $I$ and $J$ are not necessarily disjoint. A typical substrate node $v \in V$ may satisfy one or more of the following statements: (i) node $v$ is a gateway to the satellite, (ii) node $v$ is an initial demand point of the terrestrial network, (iii) node $v$ relays the traffic of other nodes, to one or more of the gateways. The ideal solution will introduce a set of terrestrial nodes for gateway placement which result in the cheapest deployment \& routing, together with the corresponding routes from all the demand points to the satellite, while satisfying the design constraints.


Regarding the delay of the network we consider only propagation delay. 
The propagation delay of a path in the network is the sum of the propagation delays over its constituting links. A GEO satellite is considered in the particular system model \cite{GW}. Let $d_{uv}$ represent the contribution of the terrestrial link $(u,v)$ to the propagation delay of a path which contains that link. The propagation delay from a gateway to the satellite is constant \cite{GWC}. Moreover, we consider multi-path routing for the traffic demands. Therefore, we define a flow $i\rightarrow j$ as the fraction of the traffic originated at node $i \in I$, routed to the satellite through gateway $j \in J$. 

Once all the flows corresponding to node $i$ are determined, all the routing paths from node $i$ to the satellites are defined. Let $c_j$ denote the cost associated with deploying a satellite gateway at node $j$ and $c_{uv}$ the bandwidth unit cost for each link $(u,v)$. We also define $a_i$ as the traffic demand rate of node $i$ which is to be served by the satellite. The capacity of each link $(u,v)$ is denoted by $q_{uv}$ while the capacity of the gateway-satellite link is $q_j$ for gateway $j$. Table \ref{paramv} summarizes all the notations used for the parameters and variables throughout the paper.

\begin{table*}[ht]
\centering
\begin{center}
\scalebox{0.9}{
\begin{tabular}{|c|c|}
\hline
Variables & Description\\
\hline
$y_j$ & The binary decision variable of gateway placement at node $j$\\
$x_{ij}$ & The fraction of traffic demand originating at $i$ passing through gateway $j$\\
$f_{uv}^{ij}$ & The amount of traffic originating at $i$ assigned to gateway $j$ passing through the link $(u,v)$\\
\hline
Parameters & Description\\
\hline
$G= (V,E)$ & Terrestrial network graph\\
$J$ & The set of potential nodes for gateway placement \\
$I$ & The set of demand points\\
$c_j$ & The cost associated with deploying a satellite gateway at node $j$\\
$c_{uv}$ & Bandwidth unit cost for link $(u,v)$\\
$a_i$ & Traffic demand rate of node $i$\\
$q_{uv}$ & Capacity of link $(u,v)$\\
$q_j$ & Capacity of gateway-satellite link for gateway $j$\\
\hline
\end{tabular}}
\caption{System model parameters and variables}
\label{paramv}
\end{center}
\end{table*}


Since the propagation latency is usually the dominant factor in determining the network delay \cite{taghlid}, we first develop a joint satellite gateway placement and routing (JSGPR) MILP formulation to minimize the cumulative gateway placement cost with hard constraints on the average 
propagation delay for the traffic of each demand point.


Following that, we will derive a variant of JSGPR with load balancing (JSGPR-LB) which aims at mutually optimizing the overall cost and the load assigned to all the deployed gateways. Finally, we use an LP-based approximation approach to reduce the time-complexity of the proposed scheme at the cost of a sub-optimal gateway placement.


\section{Problem Formulation}
\label{sec:problem}

\subsection{JSGPR MILP Formulation I}
Inspired by \cite{Bell}, we formulate a baseline for the JSGPR problem as the capacitated facility location-routing problem considering: 
\begin{itemize}
    \item The set of binary variables $\textbf{y}$, where $y_j$ expresses the placement of a gateway at node $j$.
    \item The set of continuous variables $\textbf{x}$, where $x_{ij}$ expresses the fraction of traffic demand originating at $i$ passing through gateway $j$.
    \item The set of continuous variables $\textbf{f}$, where $f_{uv}^{ij}$ expresses the amount of traffic demand originating at $i$, assigned to gateway $j$ passing through the link $(u,v)$.
\end{itemize}
 \vspace{2mm}
 We note that for a gateway $u$, the variable $f_{uu}^{iu}$ represents the amount of traffic which is originated at node $i$ and is forwarded to the satellite through the gateway placed at node $u$. 

The resulting MILP formulation is as follows:
 \begin{align}
     \textbf{Minimize} \quad \sum_{j\in J}c_j {y_j} + 
     \sum_{i \in I}\sum_{j \in J}\sum_{(u,v) \in E}c_{uv}f^{ij}_{uv}
 \end{align}
 \noindent\textbf{Demand Constraints:}
 \begin{align}
     \sum_{j \in J}{x_{ij}} = 1,\; \quad \forall i \in I
     \label{c1b}
 \end{align}
 \noindent\textbf{Feasibility Constraints:}
 \begin{align}
     x_{ij}\leq y_j, \; \quad \forall i\in I, j \in J
     \label{c2b}
 \end{align}
  \noindent\textbf{Capacity Constraints:}
  \begin{align}
     \sum_{i \in I}a_i {x_{ij}} \leq q_j y_j,\; \quad \forall j \in J
     \label{c4b}
 \end{align}
 \begin{align}
     \sum_{i \in I} \sum_{j \in J} f_{uv}^{ij}  \leq q_{uv}\; \quad \forall (u,v) \in E
     \label{c5b}
 \end{align}
  \noindent\textbf{Flow Constraints: }
 \begin{align}
        \quad a_i x_{ii} + \sum_{v\in V: (i,v) \in E}\sum_{j \in J }f^{ij}_{iv} = a_i,\; \quad \forall i \in I
    \label{flow1b}
\end{align}

\begin{equation}
    \begin{aligned}
        &\quad \sum_{v \in V: (v,u) \in E}f^{ij}_{vu} - \sum_{v \in V: (u,v) \in E}f^{ij}_{uv} = a_i x_{iu} \; \\
& \quad \forall i \in I, j \in J, u \in V,  u \neq i
    \end{aligned}
    \label{flow2b}
\end{equation}

\noindent\textbf{Domain Constraints: }
\begin{align}
    \quad y_{j} \in \{0, 1\},\; \quad \forall j \in J
    \label{dom1b}
\end{align}
\begin{align}
    \quad  x_{ij} \in [0, 1],\; \quad \forall i \in I, j \in J
    \label{dom2b}
\end{align}
\begin{align}
    \quad f_{uv}^{ij} \geq 0,\; \quad \forall (u,v) \in E, i \in I, j \in J
    \label{dom3b}
\end{align}

The objective function minimizes the total cost comprised of two terms; the first one represents the cost of installing and operating a gateway at location $j$, aggregated over the total number of gateways used. 
The second term corresponds to the transport/connection cost from the demand originating at $i$ to its assigned gateway $j$, aggregated for all demands.

Constraints set \eqref{c1b} assures that traffic demands are supported by the selected gateways. Feasibility constraints set \eqref{c2b} makes sure that demands are only served by open gateways.
Constraints \eqref{flow1b} and \eqref{flow2b} enforce the flow conservation.
The domains of $y_i$, $x_{ij}$ and $f_{uv}^{ij}$ variables are defined in constraints \eqref{dom1b}, \eqref{dom2b}, \eqref{dom3b}, respectively. 

In the aforementioned formulation, 
 we can express the set of $x_{ij}$ variables, utilizing the corresponding last-hop flow variables:

\begin{equation}
    x_{ij} = \frac{\sum_{u \in V}{f^{ij}_{uj}}}{a_i},\; \quad \forall i \in I, j \in J
    \label{aux}
\end{equation}

We can replace the set of $x_{ij}$ variables as in \eqref{aux}. The new MILP model for JSGPR is presented in the next subsection.


\subsection{JSGPR MILP Formulation II}

The new MILP formulation is as follows:

\vspace{2mm}
\begin{equation}
    \textbf{Minimize} \quad
    \sum_{j \in J}c_jy_{j} + \phi\sum_{i \in I}\sum_{j \in J}\sum_{(u,v) \in E}c_{uv}f^{ij}_{uv} \;
    \label{obj}
\end{equation}

\noindent\textbf{Demand  Constraints: }
\begin{equation}
    \quad\sum_{j \in J}\sum_{u \in V} f^{ij}_{uj} = a_i,\; \quad \forall i \in I 
    \label{demgar}
\end{equation}

\noindent\textbf{Feasibility  Constraints: }
\begin{equation}
    \quad \sum_{u \in V} f^{ij}_{uj} \leq y_j a_i,\; \quad \forall i \in I, \forall j \in J 
    \label{leq}
\end{equation}




\noindent\textbf{Capacity Constraints: }
  \begin{equation}
     \sum_{i\in I}\sum_{u\in V}{f^{ij}_{uj}} \leq q_j y_j,\; \quad \forall j \in J
     \label{c4b2}
 \end{equation}
\begin{equation}
    \quad \sum_{i \in I}\sum_{j \in J}f^{ij}_{uv} \leq q_{uv},\; \quad \forall (u,v) \in E
    \label{linkcap}
\end{equation}

\noindent\textbf{Flow Constraints: }
\begin{equation}
         \sum_{u \in V: (u,i) \in E}{f^{ii}_{ui}} + \sum_{v\in V: (i,v) \in E}\sum_{j\neq i \in J }f^{ij}_{iv} = a_i,\; \quad \forall i \in I
    \label{flow1}
\end{equation}

\begin{equation}
    \begin{aligned}
        &\quad \sum_{v \in V: (u,v) \in E}f^{ij}_{vu} - \sum_{v \in V: (u,v) \in E}f^{ij}_{uv} = \sum_{u \in V}{f^{iu}_{uu}} \; \\ 
& \quad \forall i \in I, j \in J, u \in V,  u \neq i
    \end{aligned}
    \label{flow2}
\end{equation}


\noindent\textbf{Domain Constraints: }
\begin{equation}
    \quad y_{j} \in \{0, 1\},\; \quad \forall j \in J
    \label{dom1}
\end{equation}

\begin{equation}
    \quad f_{uv}^{ij} \geq 0,\; \quad \forall (u,v) \in E, i \in I, j \in J
    \label{dom2}
\end{equation}

\vspace{2mm}

Where $\phi = \frac{1}{(\sum_{i\in I}{a_i})}$ is the normalization factor between the two terms of the objective function.
Additionally, in order to meet the average delay requirement for each demand point $i$ we will impose a new constraint exploiting the corresponding flow variables: 

\begin{equation}
    \quad \sum_{j \in J}\sum_{(u,v) \in E}\frac{f^{ij}_{uv}}{a_i}d_{uv}\leq d_{max} \; \quad \forall{i \in I}
    \label{newdelgar}
\end{equation}
where $d_{max}$ is the maximum allowed average delay for each terrestrial node. We note that we have ignored the delay of the terrestrial-satellite link in the above calculation, since it is a constant term as explained in section \ref{sec:desc}.




\subsection{JSGPR-LB MILP Formulation}

 In the aforementioned specification, the assigned load to a gateway is solely bounded by the capacity constraints \eqref{c4b2} and \eqref{linkcap}. We define a new single decision variable $l_{max}$ to represent the maximum traffic assigned to a gateway, common for all the selected gateways. We add $l_{max}$ as an additional term to the objective function. The new objective function is described in \eqref{objlm}:

\begin{align}
       \textbf{Minimize} \quad(\sum_{j \in J}c_jy_{j}+\phi\sum_{i \in I}\sum_{j \in J}\sum_{(u,v) \in E}c_{uv}f^{ij}_{uv}) + \alpha l_{max}\;
    \label{objlm}
\end{align}
where $\alpha$ is a constant factor determining the balance between the two terms of the objective function.  
Also, we change  constraints \eqref{c4b2} as following:
\begin{equation}
         \quad \sum_{i \in I}\sum_{u \in V} f^{ij}_{uj} \leq l_{max}\; \quad \forall j \in J 
\end{equation}
where $l_{max} \leq q_j \quad \forall j \in J$. We will call this last model, $JSGPR-LB$. The resulted optimization problem aims to balance the load of the gateways in conjunction with the cost of the gateway deployment \cite{manet}, \cite{lb}. 

\subsection{LP Relaxation and Approximation Algorithm}

Since the MILP model is known to be NP-hard, the problem is intractable for larger scale networks \cite{Bell}. For the aforementioned JSGPR and JSGPR-LB MILP formulations, the optimal fractional solution is computed for the problem via linear programming relaxation. The relaxed problem can be solved by any suitable linear programming method, in polynomial time. A rounding technique is applied, similar to \cite{torkzaban2019trust} to obtain the integer solution of the MIP problem. The resulting multi-commodity flow allocation problem is solved to identify the routing paths. 



\section{Performance Evaluation}
\label{sec:evaluation}
In this section we evaluate the performance of our satellite gateway placement method. We first describe the simulation environment setup and scenarios, then we review the performance evaluation results. 
\subsection{Performance Evaluation Setup}
\label{evaluation:setup}
We evaluate the performance of our JSGPR and JSGPR-LB approaches on multiple real network topologies publicly available at the Topology Zoo \cite{zoo}. The five different topologies we consider are listed in table \ref{topo}. The link lengths and capacities are extracted from the topology zoo. The propagation delays are calculated based on the 
lengths of the links with the propagation velocity of $C = 2 \times 10 ^8 m/s$ \cite{speed}.  The value of $d_{max}$ is set to $10 ms$, and the deployment cost for each node is taken from a uniform random generator $~U(500, 1000)$. The unit bandwidth cost for all the links is set to be equal to $1$. Also, the value of $q_j$ is set to $240 Mbps$ for all the gateways.
To develop and solve our MILP and LP models we use the CPLEX commercial solver. Our tests are carried out on a server with an Intel i5 CPU at 2.3 GHz and 8 GB of main memory.


\begin{table}[b]
\centering
\begin{tabular}{||c c c||} 
 \hline
 Topology & Nodes & Links  \\ 
 \hline\hline
 Sinet & 13 & 18  \\ 
 Ans & 18 & 25  \\
 Agis & 25 & 32  \\
 Digex & 31 & 35  \\
 Bell Canada & 48 & 64 \\ 
 \hline
\end{tabular}
\caption{Summary of the studied topologies}
\label{topo}
\end{table}

\subsection{Evaluation Scenarios}
\label{evaluation:setup}
We have conducted two sets of experiments for two different cases. In the first case we benchmark our LP-based approximation algorithm against the MILP-based optimal one for JSGPR, whereas the second case aims to observe the impact of our load balancing approach on the gateway placement. 
For both cases and for all topologies, if $c^i_{max}$ is the maximum capacity among all the outgoing links of node $i$, the traffic rate originated at node $i$ is taken from a uniform distribution $~ U(\frac{2c^i_{max}}{3}, c^i_{max})$. Specifically, we will use the following metrics to evaluate the performance of our algorithm:

\begin{itemize}
\item \textbf{Solver Runtime} is the amount of time the CPLEX solver takes to solve the generated MILP or LP instances.  
\item \textbf{Average Delay} as explained in section \ref{sec:desc}.
\item \textbf{Total Cost} is the total cost of deploying the satellite gateways and routing the traffic.
\item \textbf{Gateway Load} is the amount of traffic assigned to the gateways after the gateway placement problem is solved.
\end{itemize}

\subsection{Evaluation Results}
\label{sec3}
\begin{figure*}[t]
\begin{center}
\begin{minipage}[h]{0.225\textwidth}
\includegraphics[width=1\linewidth]{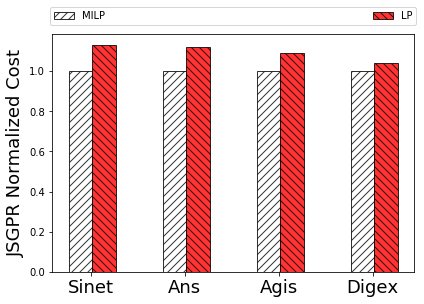}
\caption{Exp-A: Normalized Total Cost}
\label{fig:acc}
\end{minipage}
\hspace{1em}
\begin{minipage}[h]{0.225\textwidth}
\includegraphics[width=1\linewidth]{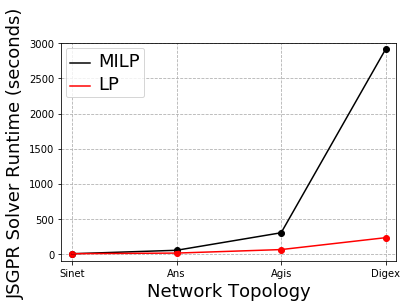}
\caption{Exp-A: Average Solver Runtime}
\label{fig:util}
\end{minipage}
\hspace{1em}
\begin{minipage}[h]{0.225\textwidth}
\includegraphics[width=1\linewidth]{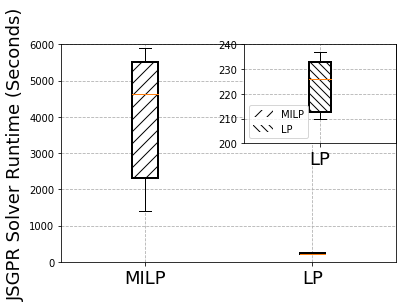}
\caption{Exp-A: Solver Runtime for Digex}
\label{fig:prev}
\end{minipage}
\hspace{1em}


\begin{minipage}[h]{0.225\textwidth}
\includegraphics[width=1\linewidth]{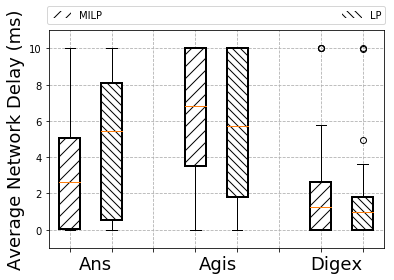}
\caption{Exp-A: Average Delay ($d_{max} = 10 ms$ )}
\label{fig:agg}
\end{minipage}
\hspace{1em}
\begin{minipage}[h]{0.225\textwidth}
\includegraphics[width=1\linewidth]{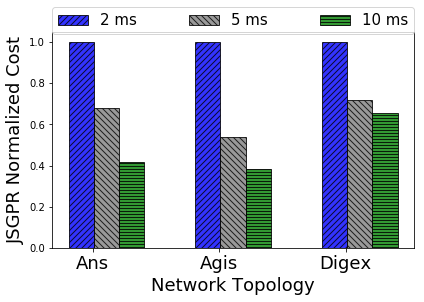}
\caption{Exp-A: Total Cost with Varying Delay Bound}
\label{fig:comp}
\end{minipage}
\hspace{1em}
\begin{minipage}[h]{0.225\textwidth}
\includegraphics[width=1\linewidth]{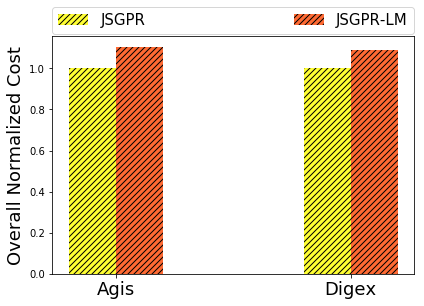}
\caption{Exp-B: Normalized Total Cost}
\label{fig:cost2}
\end{minipage}
\hspace{1em}
\begin{minipage}[h]{0.225\textwidth}
\includegraphics[width=1\linewidth]{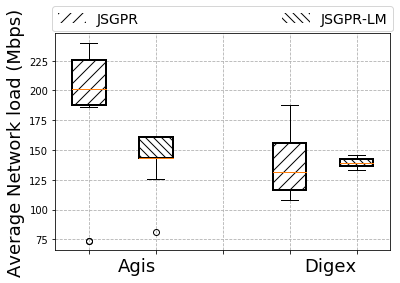}
\caption{Exp-B: Gateway Load }
\label{fig:load_profile}
\end{minipage}
\hspace{1em}

\end{center}
\end{figure*}

\subsubsection{Experiment A - Approximation vs. Exact Method}Fig. \ref{fig:acc} reflects the average normalized total cost for both the optimal MILP-based approach and the LP-based approximation of JSGPR. Due to the suboptimal placement, the LP-based approach results in additional deployment costs within the range of at most $13\%$ of the optimal placement, but as the scale of the network gets larger this gap decreases. For Digex the approximate approach leads to about $4\%$ increase in the deployment cost. 

Fig. \ref{fig:util} depicts the average runtime for both the MILP and LP formulations for the $4$ topologies. We note that for larger networks, CPLEX failed to provide the solution for the MILP problem, while our LP model continued to solve the problem within the expected time limit. Particularly, for the Bell-Canada topology, in $40\%$ of the runs, CPLEX was not able to find any feasible solutions within the first $10$ hours while the approach via approximation could provide the suboptimal placement in less than $15$ minutes. The solver runtime for Digex is shown in \ref{fig:prev}. The average runtime for the LP is $230$ seconds while for the MILP, it is around $3000$ seconds.

Fig. \ref{fig:agg}, represents the average delay for Ans, Agis, and Digex. The average of the expected experienced delay under the MILP model is $2.95, 6.18,$ and $ 2.15$ seconds, while this value under the LP model is $4.68, 5.6,$ and $1.65$ seconds. We note that the suboptimal procedure leads to additional deployed gateways in Agis and Digex, which in return will have the nodes end up closer to the gateways and consequently, experience lower average delay.
Further, an insightful observation is that, in Agis, $d_{max}$ is relatively a tight upper bound on the average delay experienced by each node. Therefore, the delay profile is pushed towards its upper bound, whereas, in Digex, due to the low density of links, more gateways are required to be deployed on the terrestrial nodes, which will make the gateways available more quickly; Therefore, the delay profile is inclined towards its lower bound.    

Fig. \ref{fig:comp} depicts the normalized cost of the JSGPR problem for different values of $d_{max}$. For instance, as indicated by this figure, if in Digex, a delay of $10 ms$ is tolerable instead of $2ms$, a $35\%$ reduction in cost results; Similarly, upgrading the service from a delay of $5ms$ to $2ms$ in Agis, will almost double the cost.

Overall, the aforementioned figures illustrate that the performance of our approximation algorithm is very close to the exact approach, but with an important advantage of reduced time complexity which shows the efficiency of our proposed approximation method.

\subsubsection{Experiment B - The Impact of Load Minimization}
Fig. \ref{fig:cost2} shows the profile of the average load on the gateways for JSGPR, 
and JSGPR-LB, considering the Agis and Digex topologies. As expected, JSGPR-LB is more costly, since in order to evenly share the load between the gateways, 
a larger number of gateways will be required. 

Fig. \ref{fig:load_profile} depicts the profile of the load assigned to the gateways. In both depicted topologies, the load profile under JSGPR-LB is very thin and concentrated over its average, proving the efficiency of the formulation. 
Lower load on the placed gateways is achieved by the sub-optimal placement of the gateways (due to the larger number of gateways) which results in a more expensive gateway placement. As depicted in Fig. \ref{fig:cost2} the total cost of gateway placement in the studied topologies 
for JSGPR-LB is above and within the range of $16\%$ of the optimal placement cost achieved by JSGPR.
\section{Related Work}
\label{sec:relatedwork}
 Although the gateway placement problem over a general network is very-well studied, the satellite gateway placement problem on an ISTN is fairly new. 
 For sensor \cite{bzoor}, vehicular\cite{mezza}, wireless mesh \cite{gorbe} and MANETs \cite{manet}, different approaches have been proposed with different optimization objectives such as load balancing \cite{lb}, latency minimization \cite{delay}, reliability maximization \cite{rel}, etc.

 The recent works on ISTNs , mostly focus on optimizing the average network latency\cite{GW} or reliability\cite{GWC}, \cite{capac}, \cite{taghlid}, \cite{same}. In \cite{GW}, the authors propose a particle swarm-based optimization approximation (PSOA) approach for mimizing the average network latency and benchmark it against an optimal brute force algorithm (OBFA) to show the imporvement in the time complexity. In \cite{GWC}, a hybrid simulated annealing and clustering algorithm (SACA) is used to provide near optimal solution for the joint SDN controller and satellite gateway placement with the objective of network reliability maximization.  For the same purpose as \cite{GWC}, authors in \cite{taghlid}, provide a simulated annealing  partion-based k-means (SAPKM) approach and compare the advantages and weaknesses of their approach with that of \cite{GWC}. In \cite{capac}, the authors consider the satellite link capacity as a limiting constraint and solve the placement problem to maximize the reliability. 
 
 There is another line of research which aims to place the satellite gateways on the aerial platforms. More precisely in \cite{cross}, a new aerial layer is added in between the terrestrial and the satellite layer which relays the traffic between the satellite and the terrestrial nodes. The authors have used a greedy optimization approach to make the gateway selection.
 
 Although the above approaches provide a great insight into the satellite gateway placement problem, they assume the number of gateways is known prior to design. Secondly, none of the above studies investigates the traffic routes from the terrestrial switches to the satellite, whereas we propose the joint optimization of satellite gateway placement together with the corresponding traffic routing for ISTNs. 
\section{Conclusions}
\label{sec:conclusions}

In this paper, we introduce the joint satellite gateway placement and routing problem over an ISTN, for facilitating the terrestrial-satellite communications while adhering to propagation latency requirements, in a cost-optimal manner. 
We also balance the corresponding load between selected gateways.To yield a polynomial solution time, we relax the integer variables and derive an LP-based rounding approximation for our model. 

In SDN-enabled ISTNs, the problem of controller placement needs to be addressed mutually with the placement of the satellite-gateways having in mind different design strategies such as cross-layer data delivery, load balancing, reliability and latency optimization, etc. Moreover, instead of placing the satellite gateways on the terrestrial nodes, aerial platforms can be an alternative choice leading to a cross-layer network design problem. The flow routing in this setting is more challenging than the presented work. These two problems are the topics of our future research.

\bibliographystyle{IEEE}
\bibliography{bibliography}

\end{document}